\documentclass[aps,pra,reprint,amsmath,superscriptaddress,floatfix]{revtex4-2}
\usepackage{amsmath}
\usepackage{amssymb}
\usepackage{graphicx}

\makeatletter
\usepackage{color}
\usepackage{xspace}

\makeatother

\begin{document}
\title{Intermediate scattering potential strength in electron - irradiated\\
 $\text{YBa}_{2}\text{Cu}_{3}\text{O}_{7-\delta}$ from London penetration
depth measurements }
\author{Kyuil Cho}
\affiliation{Ames Laboratory, Ames, IA 50011, USA}
\affiliation{Department of Physics \& Astronomy, Iowa State University, Ames, IA
50011, USA}
\author{M. Ko\'{n}czykowski}
\affiliation{Laboratoire des Solides Irradies, École Polytechnique, CNRS, CEA,
Institut Polytechnique de Paris, F-91128 Palaiseau, France}
\author{S. Teknowijoyo}
\affiliation{Ames Laboratory, Ames, IA 50011, USA}
\affiliation{Department of Physics \& Astronomy, Iowa State University, Ames, IA
50011, USA}
\author{S. Ghimire}
\affiliation{Ames Laboratory, Ames, IA 50011, USA}
\affiliation{Department of Physics \& Astronomy, Iowa State University, Ames, IA
50011, USA}
\author{M. A. Tanatar}
\affiliation{Ames Laboratory, Ames, IA 50011, USA}
\affiliation{Department of Physics \& Astronomy, Iowa State University, Ames, IA
50011, USA}
\author{Vivek Mishra}
\affiliation{Kavli Institute for Theoretical Sciences, University of Chinese Academy
of Sciences, Beijing 100190, China}
\author{R. Prozorov}
\affiliation{Ames Laboratory, Ames, IA 50011, USA}
\affiliation{Department of Physics \& Astronomy, Iowa State University, Ames, IA
50011, USA}

\date{18 November 2021}

\begin{abstract}
Temperature-dependent London penetration depth, $\lambda(T)$, of
a high quality optimally-doped $\text{YBa}_{2}\text{Cu}_{3}\text{O}_{7-\delta}$
single crystal was measured using tunnel-diode resonator. Controlled
artificial disorder was induced at low-temperature of 20~K by 2.5
MeV electron irradiation at accumulating large doses of $3.8\times10^{19}$
and $5.3\times10^{19}$ electrons per $\textrm{cm}^{2}$. The irradiation
caused significant suppression of the superconductor's critical temperature,
$T_{c}$, from 94.6 K to 90.0 K, and then to 78.7 K, respectively.
The low-temperature behavior of $\lambda\left(T\right)$ evolves from
a $T-$linear in pristine state to a $T^{2}-$behavior after the irradiation,
expected for a line-nodal $d-$wave superconductor. However, the original
theory that explained such behavior had assumed a unitary limit of
the scattering potential, whereas usually in normal metals and semiconductors,
Born scattering is sufficient to describe the experiment. To estimate
the scattering potential strength, we calculated the normalized superfluid
density, $\rho_{s}\left(t=T/T_{c}\right)=\lambda^{2}\left(0\right)/\lambda^{2}\left(t\right)$,
varying the amount and the strength of non-magnetic scattering using
a self-consistent $t-$matrix theory. Fitting the obtained curves
to a power-law, $\rho_{s}=1-Rt^{n}$, and to a polynomial, $\rho_{s}=1-At-Bt^{2}$,
and comparing the coefficients $n$ in one set, and $A$ and $B$
in another with the experimental values, we estimate the phase shift
to be around 70$^{\circ}$ and 65$^{\circ}$, respectively. We correlate
this result with the evolution of the density of states with non-magnetic
disorder. 
\end{abstract}
\maketitle

\section{Introduction}

While the microscopic mechanism of superconductivity is still actively
debated three decades after the discovery \citep{Bednorz1986}, it
is universally accepted that at least optimally-doped high$-T_{c}$
cuprates have nodal, $d-$wave, symmetry of the order parameter \citep{Xu1995,vanHarlingenRMP1995,Annett1996,SHEN200814,HighTc2001}.
Line nodes in the gap function result in linear in energy density of states
which leads to a $T-$linear variation of the superfluid density at
low temperatures, where ``low'', roughly $T_{c}/3$, is defined
as the temperature below which the superconducting order parameter
amplitude is nearly constant, thus the density of nodal quasiparticles
is determined solely by the angular variation of the superconducting
gap on the Fermi surface. Indeed, measured linear temperature dependence
of $\lambda\left(T\right)$ was among the first definitive arguments
in favor of a $d-$wave superconductivity in $\text{YBa}_{2}\text{Cu}_{3}\text{O}_{7-\delta}$
(YBCO) \citep{Hardy1993PRL,Schachinger2003}.

Another consequence of anisotropic (and, in extreme cases, nodal)
order parameter is the violation of the Anderson theorem \citep{Anderson1959a},
satisfied only for isotropic gap and non-magnetic scatterers. Abrikosov
and Gor'kov showed that spin-flip scattering on magnetic impurities
suppresses even isotropic superconducting gap \citep{AbrikosovGorkov1960ZETF}.
In case of anisotropic superconducting gap, the order parameter (hence
$T_{c}$) is suppressed by both magnetic and non-magnetic impurities
\citep{Hirschfeld1993PRB,Openov2004,Openov1997,Golubov1997} and this
can be readily extended to multiband superconductors \citep{Kogan2009PRB-2,Efremov2011}.
This universal suppression of $T_{c}$ by all types of disorder was,
indeed, observed in YBCO \citep{Giapintzakis1995,Rullier-Albenque2003PRL_YBCO_e-irr,Rullier_Albenque2000EPL_YBCO_e-irr}.

As far as low-temperature variation of $\Delta\lambda\left(T\right)$
is concerned, already early experiments showed a crossover from $T-$linear
to a $T^{2}$ variation of $\lambda\left(T\right)$, for example in
Zn-doped YBa$_{2}$Cu$_{3}$O$_{6.95}$ \citep{AchkirHardy1993PRB,BonnHardy1994PRB_YBCO_impurity}.
These results were explained by Hirschfeld and Goldenfeld \citep{Hirschfeld1993PRB},
but in order to provide a quantitative agreement with the experiment
they had to postulate unitary limit of the impurity scattering. Since
usually the opposite, weak scattering (Born) limit explains the properties
of normal metals, this feature of their theory was puzzling and attracted
significant experimental and theoretical follow up \citep{LeeHone2017}.
Analysis of the optical spectroscopy data led to a conclusion that
the intermediate (between Born and unitary) scattering phase shifts
describe YBCO \citep{Schachinger2003}. We note that majority of works
studying disorder in superconductors focuses on the suppression of
$T_{c}$ and there is little known about low-temperature thermodynamics.
During the last decade we studied both, the variation of $T_{c}$
and the changes in $\lambda\left(T\right)$, focusing on iron-based
superconductors \citep{Prozorov2006SST,ProzorovKogan2011RPP} and
here we conduct a similar study of $\text{YBa}_{2}\text{Cu}_{3}\text{O}_{7-\delta}$.

Experimentally, it is not trivial to introduce non-magnetic point
like disorder. Chemical substitution perturbs the stoichiometry and
may change the electronic band-structure. One of the clean ways to
do this is to use particle irradiation, which was used for metals
extensively since the mid of the last century \citep{Damask1963,THOMPSON1969}.
Among all projectiles, relativistic electrons provide just enough
energy to induce vacancy-interstitial Frenkel pairs. If the irradiation
is conducted at low temperatures, the recombination and clustering
are inhibited and upon warming, the interstitials migrate to various
sinks (dislocations, twin boundaries, surfaces) leaving behind a quasi
equilibrium population of vacancies that act as point-like scattering
centers. A good check of the metastable nature is the (sometimes complete)
recovery of properties upon annealing signaling that the irradiation
did not cause an irreversible damage. More details on using electron
irradiation for superconductors can be found elsewhere \citep{Cho2018SST_review}.
In YBCO, the effect of electron irradiation on transport properties
and transition temperature is described in Refs.\citep{Giapintzakis1995,Rullier-Albenque2003PRL_YBCO_e-irr,Rullier_Albenque2000EPL_YBCO_e-irr}.

We used relatively large doses of electron irradiation to induce non-magnetic
point-like disorder in high quality optimally doped $\text{YBa}_{2}\text{Cu}_{3}\text{O}_{7-\delta}$
single crystal. The measured suppression of the superconducting transition
temperature, $T_{c}$, gives the estimate of the dimensionless scattering
rate $\Gamma/T_{c0}$ (see appendix), whereas the variation of the
low-temperature London penetration depth, $\Delta\lambda\left(t\right)$,
is related directly to the normalized superfluid density, $\rho_{s}\left(t\right)\thickapprox1-2\Delta\lambda\left(t\right)/\lambda\left(0\right),$whose
temperature dependence is sensitive to the structure of the superconducting
order parameter \citep{Prozorov2006SST,ProzorovKogan2011RPP}. Here,
$t=T/T_{c}$ is the (scattering - dependent) reduced temperature,
and $\lambda\left(0\right)$ is the magnitude of the London penetration
depth at $T=0$. The impurity potential is parameterized in terms
of the scattering phase shift, where very small values correspond
to weak scattering centers (Born limit) and a phase shift close to
$\theta=90^{\circ}$ represents unitary scatterers. (Note that in
figures we show $\theta=0^{\circ}$ for Born limit for convenience.
Of course, there is a small phase shift even in that regime.) We used
$t-$matrix theory to calculate the superfluid density and then fitted
the experimental and numerical data to the same models. This leads
to the estimates the phase shift in YBCO at $\theta=65^{\circ}-70^{\circ}$,
which agrees with the theoretical conclusion that the phase shift
is intermediate between Born and unitary limits \citep{Schachinger2003}.
These values correspond to a regime where the density of states close
to zero energy become sensitive to scattering. 

\section{Experimental}

Single crystals $\text{YBa}_{2}\text{Cu}_{3}\text{O}_{7-\delta}$
were grown in yttria stabilized zirconia crucibles with subsequent
annealing to achieve the highest transition temperature as described
elsewhere \citep{Kaiser1987}. Similar crystals were used in multiple
studies with the variety of techniques. The crystal used for the current
study has dimensions of $0.8\times0.51\times0.01\;\textrm{mm}^{3}$.

The 2.5 MeV electron irradiation was performed at the SIRIUS Pelletron
facility of the Laboratoire des Solides Irradies (LSI) at the École
Polytechnique in Palaiseau, France \citep{Cho2018SST_review,Rullier-Albenque2003PRL_YBCO_e-irr}.
The acquired irradiation dose is conveniently measured in C/cm$^{2}$,
where 1 C/cm$^{2}$ = 6.24 $\times$ 10$^{18}$ electrons/cm$^{2}$.
The same crystal was irradiated twice, first receiving a dose of 6.15
C/cm$^{2}$ ($3.8\times10^{19}$ electrons per $\textrm{cm}^{2}$)
after which the measurements were repeated, and then more irradiation
was added bringing the total cumulative dose of 8.5 C/cm$^{2}$ ($5.3\times10^{19}$
electrons per $\textrm{cm}^{2}$). Appendix B details the irradiation
cross-sections and provides more information on the damage induced.

The in-plane London penetration depth $\Delta\lambda$ (T) was measured
before and after each irradiation run using a self-oscillating tunnel-diode
resonator technique (TDR) described in detail elsewhere \citep{VanDegrift1975RSI,Prozorov2000PRB,Prozorov2000a,Prozorov2006SST,ProzorovKogan2011RPP,Carrington2011}.
The TDR circuit resonates at \textasciitilde 14 MHz and the frequency
shift is measured with precision better than one part per billion
(ppb). The inductor coil generates an AC excitation magnetic field,
$H_{ac}<20~\text{mOe}$, hence the sample is always in the Meissner
state at the temperatures of interest. In the experiment, the sample
was mounted on a 1 mm diameter sapphire rod and inserted into a 2
mm diameter inductor coil. The coil and the sample are in vacuum inside
a $^{3}\textrm{He}$ cryostat. The TDR circuit is actively stabilized
at 5 K, and the sample is controlled separately from 0.4 K up by independent
LakeShore controllers. It is straightforward to show that the change
of the resonant frequency when a sample is inserted into the coil
is proportional to the sample magnetic susceptibility as long as the
change of the total inductance is small and one can expand, $\Delta f/f_{0}\approx\Delta L/2L_{0}$
where $2\pi f_{0}=1/\sqrt{CL_{0}}$ with subscript $"0"$ referring
to an empty resonator. The coefficient of proportionality that includes
the demagnetization correction, is measured directly by pulling the
sample out of the resonator at the base temperature \citep{Prozorov2000PRB,Prozorov2021}.

\begin{figure}[tb]
\includegraphics[width=7.5cm]{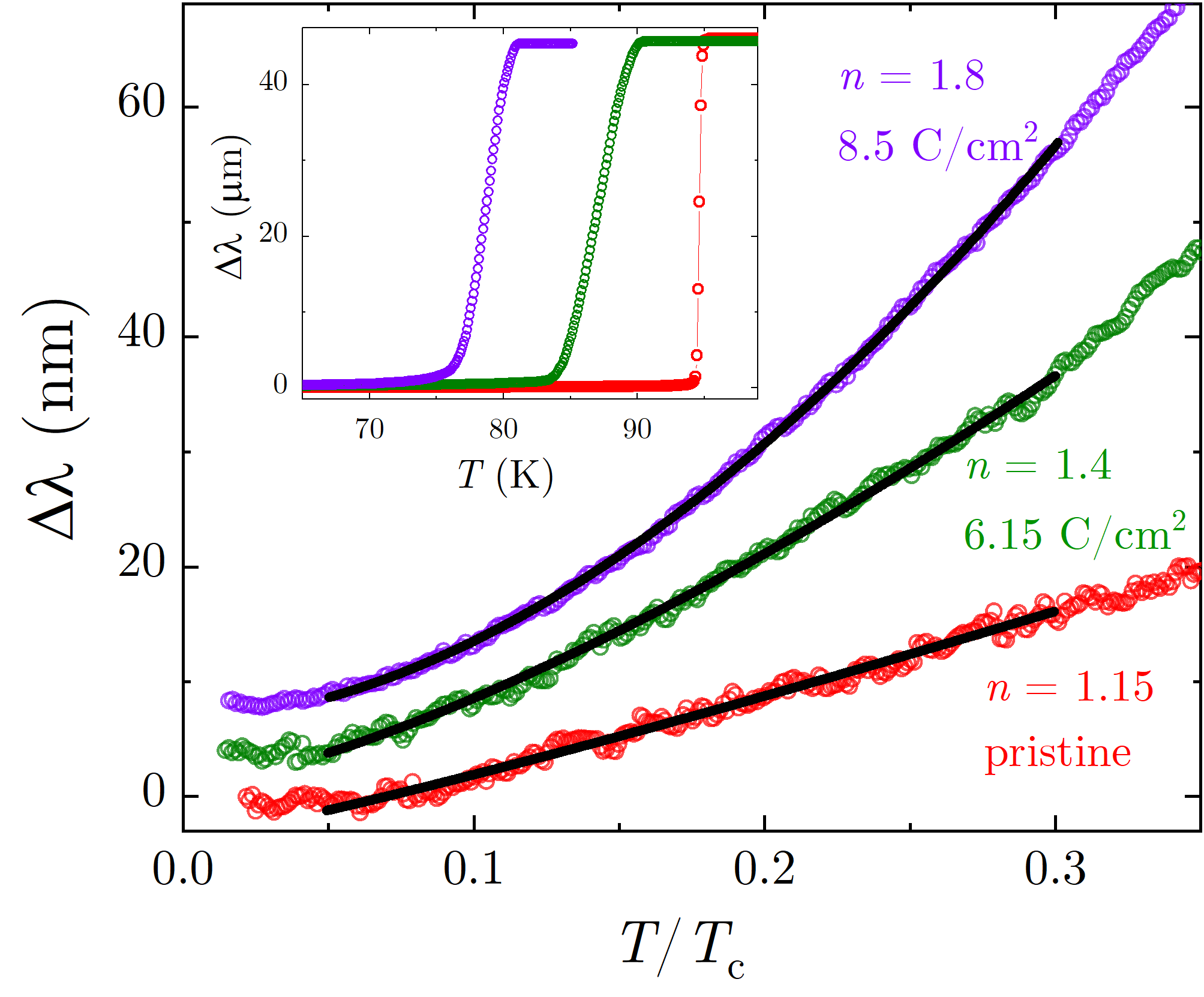} \centering \caption{Variation of the London penetration depth, $\Delta\lambda\left(t\right)$,
measured before and after the 2.5 MeV electron irradiation first with
dose of 6.15 C/cm$^{2}$ ($3.8\times10^{19}$ electrons per $\textrm{cm}^{2}$),
and then more irradiation was added bringing the total cumulative
dose of 8.5 C/cm$^{2}$ ($5.3\times10^{19}$ electrons per $\textrm{cm}^{2}$).
Main panel shows $\Delta\lambda\left(t\right)$ vs. $t=T/T_{c}$ fitted
to the power law, $\sim t^{n}$. The exponent $n$ changes from practically
linear, $n=1.15$ in pristine sample to to $n=1.8$ upon irradiation,
suggesting the crossover from $t-$linear to $t^{2}$ dependence.
The inset shows a full temperature range showing the substantial suppression
of $T_{c}$ by the irradiation. }
\label{fig1} 
\end{figure}

In this work we analyze the variation of the London penetration depth,
$\Delta\lambda=\lambda\left(T\right)-\lambda\left(T_{min}\right)\approx\lambda\left(T\right)-\lambda\left(0\right)$,
where $T_{min}=0.02T_{c}$ is the minimum temperature of the experiment.
The normalized superfluid density, $\rho_{s}\equiv\lambda^{2}\left(0\right)/\lambda^{2}\left(T\right)=1/\left(1+\Delta\lambda/\lambda\left(0\right)\right)^{2}\approx1-2\Delta\lambda/\lambda\left(0\right)$.
This ratio does not depend on $\lambda\left(0\right)$, because the
penetration depth is given roughly by $\lambda\left(t\right)/\lambda\left(0\right)\approx1/\sqrt{1-t^{p}}$
in the whole temperature interval, $0\leq t\leq1$ \citep{Prozorov2006SST}.
Here $t=T/T_{c}$ and the exponent $p=2$ or $4$ depending on the
pairing type \citep{Prozorov2006SST}. Therefore, measurements of
$\Delta\lambda\left(t\right)$ provide direct access to the structure
of the superconducting order parameter via the superfluid density,
$\rho_{s}\left(t\right)$.

\section{Results and Discussion}

Figure \ref{fig1} shows the variation of London penetration depth
in the same crystal measured before and after two exposures to 2.5
MeV electron irradiation. As shown in the inset, pristine sample exhibits
a sharp transition at 94.6 K, indicative of a high-quality optimally
doped crystal \citep{HighTc2001}. Upon irradiation, $T_{c}$ (onset)
substantially decreased, first from 94.6 K in pristine state to 90.0
K after 6.15 C/cm$^{2}$, and then to 78.7 K after a total dose of
8.5 C/cm$^{2}$ was applied. The sharp superconducting transition
in the pristine sample, $\Delta T_{c}<0.8$ K, becomes broader after
the first irradiation run, $\Delta T_{c}$ = 6.5 K at 6.15 $\textrm{C/cm}{}^{2}$,
but then becomes narrower, $\Delta T_{c}$ = 5.1 K, after 8.5 C/cm$^{2}$.
The broadening is most likely due to some initial inhomogeneity of
the produced defects density which becomes more homogeneous in time
slowly annealing at room temperature. The measurement after first
irradiation of 6.15 C/cm$^{2}$ was conducted rather quickly within
a week of the irradiation run, but the measurement after second irradiation
bringing the total dose to 8.5 C/cm$^{2}$ was made about a month
later. Perhaps, this will be interesting to study separately, but
for the present work it is important that the transition temperature
is well defined and that the region of the broader transition is far
from the low-temperature region of interest for the fitting of the
London penetration depth.

\begin{figure}[tb]
\centering \includegraphics[width=7.5cm]{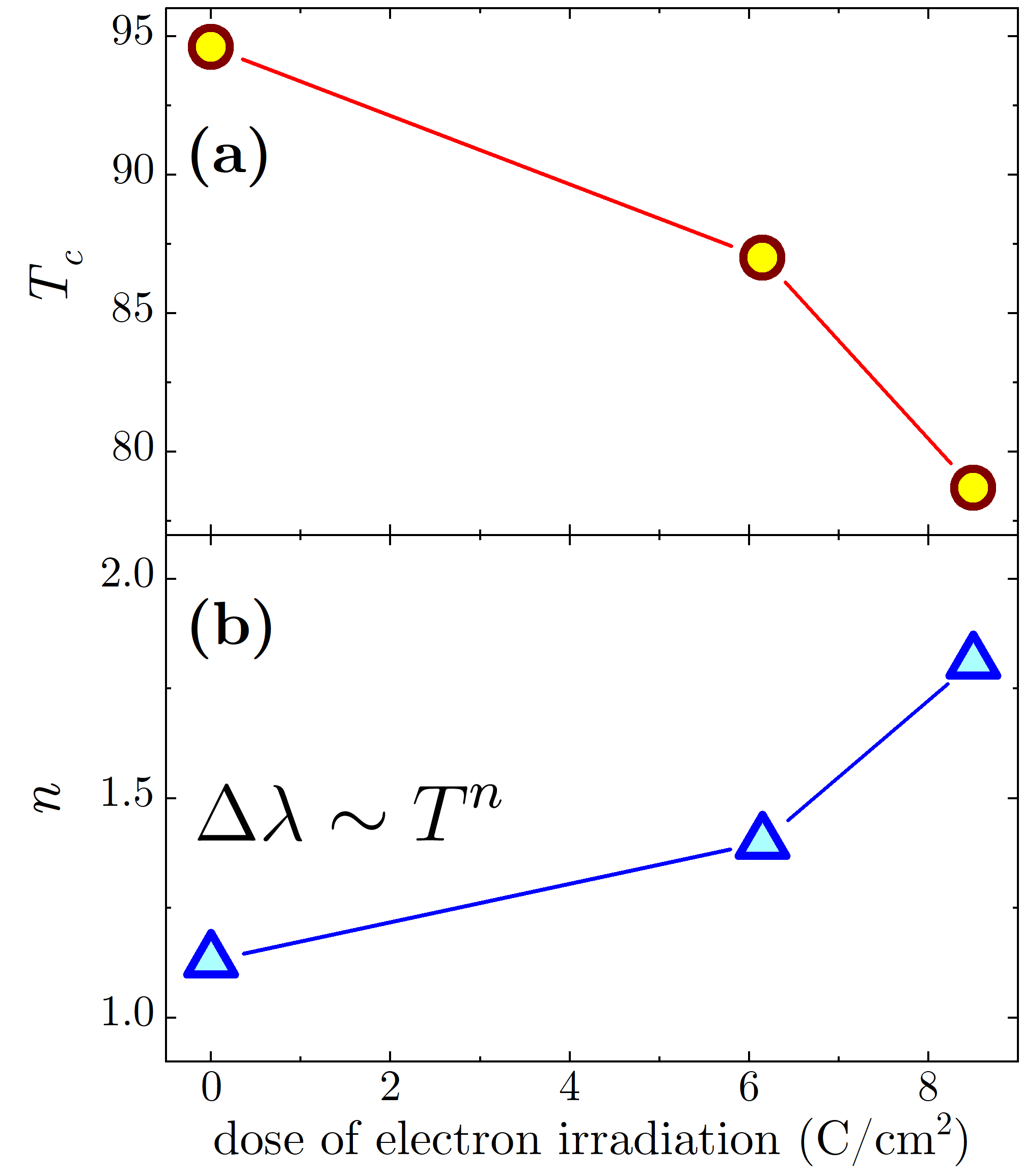} \caption{Analysis of the data presented in Fig.\ref{fig1}. (a) Superconducting
transition temperature, $T_{c}$. (b) Power-law exponent, $n$, as
a function of the cumulative dose of electron irradiation showing
a systematic evolution from a clean $T-$linear to a dirty $T^{2}$
behavior expected for a $d-$wave superconductor.}
\label{fig2} 
\end{figure}

There are two ways to analyze the differences between the curves presented
in Fig.\ref{fig1}. One is to perform a power-law fitting up to $0.3T_{c}$,
$\Delta\lambda\sim T^{n}$, which has been extensively used for the
analysis of pair-breaking scattering in $s_{\pm}$ iron-based superconductors
\citep{ProzorovKogan2011RPP,Cho2018SST_review}. Another, is to perform
a polynomial fit, $2\Delta\lambda/\lambda\left(0\right)=At+Bt^{2}$,
where the factor of 2 comes from the expansion, $\rho_{s}\approx1-2\Delta\lambda/\lambda\left(0\right)$.
The former approach has an advantage of being independent of $\lambda\left(0\right)$,
because only the exponent $n$ matters. However, using exponent $n$
as a fitting parameter is more sensitive to the choice of the fitting
range and it has a less transparent meaning for dimensional quantities.
The quadratic polynomial is a standard way to analyze such results
and it offers both coefficients $A$ and $B$, which however depend
on the choice of $\lambda\left(0\right)$. We resolve this problem
by using $t-$matrix calculations to estimate the scattering-dependent
magnitude of the London penetration depth, $\lambda(0,\Gamma)/\lambda(0,0)$
and then use the well-known clean-limit value of $\lambda(0,0)\approx150$
nm from the previous measurements of high quality YBCO single crystals
\citep{Bonn1993,Hardy1993PRL,Prozorov2000a}.

We note that the temperatures below $0.05T_{c}$ were excluded from
the analysis due to a clear change in behavior. In YBCO there are
several possible complications at very low temperatures that lead
to $1/T$ - type behavior, such as Andreev bound states observed with
our technique before \citep{Carrington2001} and/or effects of isolated
impurities predicted for a $d-$wave superconductor \citep{Tsai2002}.
However, thanks to high $T_{c}$, this is only a relatively small
part in normalized units, $t=T/T_{c}$. Figure \ref{fig2}(a) shows
the suppression of $T_{c}$ as function of irradiation dose, and Fig.\ref{fig2}(b)
shows the best-fit exponent varying from $n=1.15$ in the pristine
state to $n=1.8$ after second irradiation, clearly supporting the
expected evolution from $T-$linear to $T^{2}$ behavior with the
increase of the disorder scattering \citep{Hirschfeld1993PRB}. Since
the decrease of $T_{c}$ is a monotonic function of the scattering
rate, $\Gamma$, see Fig.\ref{figS1} of the Appendix A, it can be
used as the quantitative measure of the latter, thus eliminating the
need to evaluate $\Gamma$ explicitly.

\begin{figure}[tb]
\includegraphics[width=7.5cm]{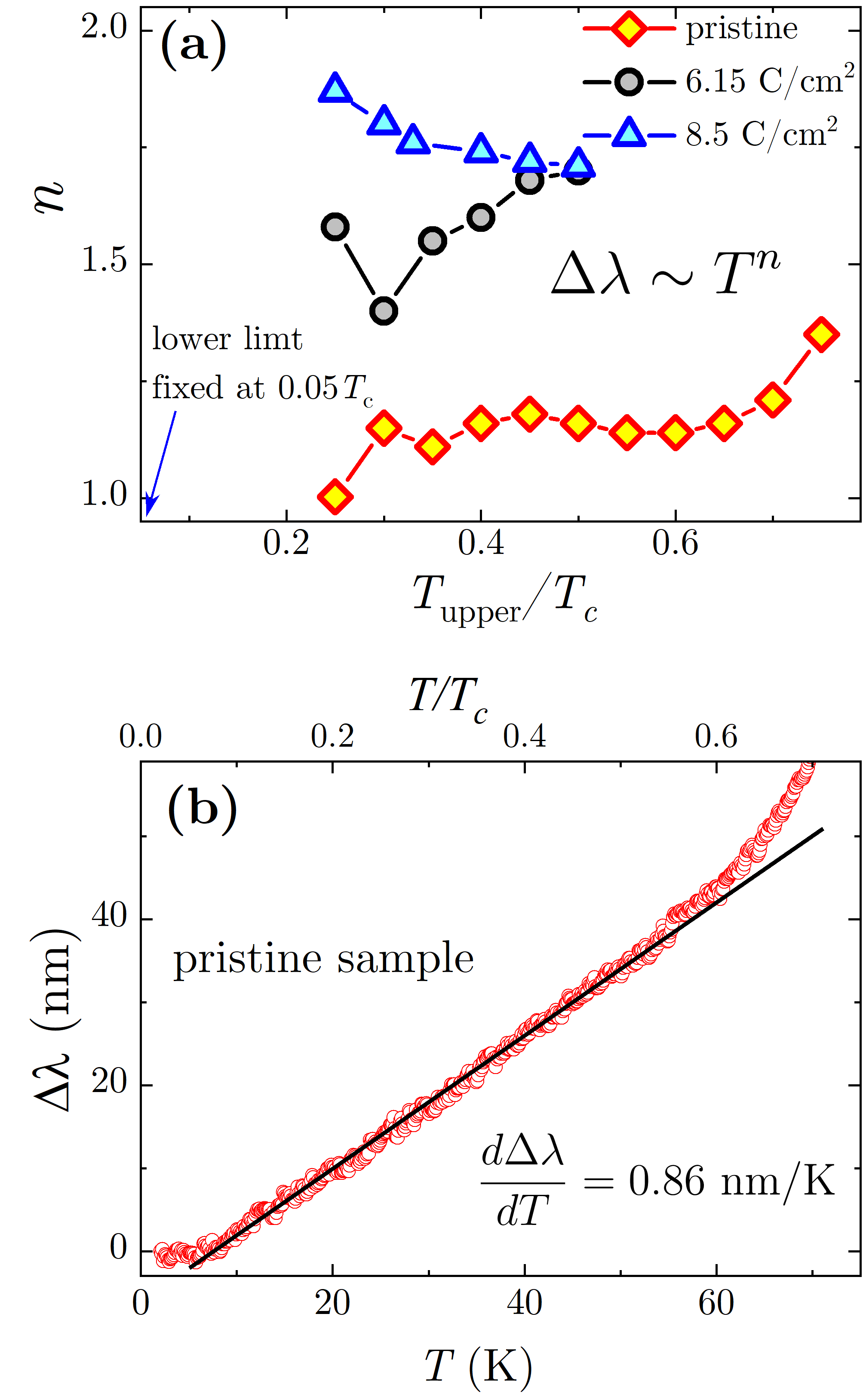} \centering \caption{\textbf{(a)} Results of a power-law fit, $\Delta\lambda\sim T^{n}$,
as a function of the upper limit of the fitting range, see text for
explanation. The lower limit was fixed at $T=0.05T_{c}$. The exponent
$n$ jumps upon irradiation from $n\approx1$ in pristine state, toward
$n=2$ in the irradiated sample, inconsistent with the Born scattering
limit. \textbf{(b)} London penetration depth in the pristine sample
exhibiting an extended range of close to $T-$linear temperature dependence,
up to 60 K (about 0.63 $T_{c}$).}
\label{fig3} 
\end{figure}

The power-law fit is explored in Fig.\ref{fig3} where the top panel
(a) summarizes the exponent $n$ obtained from the fit using different
temperature intervals with the indicated upper limit for all three
sample states considered here. This is done to explore how robust
the exponent $n$ is. Figure \ref{fig3}(b) shows the linear temperature
dependence of the London penetration depth in a pristine state, behaving
as expected for a clean $d-$wave superconductor with line nodes \citep{Xu1995,Annett1996}.
Interestingly, the linearity is extended up to quite high temperatures
around $60$ K ($0.63T_{c}$). The abrupt change below $0.05T_{c}$
is clearly not a smooth evolution and is likely where $1/T$ physics,
unrelated to scattering, takes over. For the irradiated state, the
situation is drastically different. Whereas the exponent, $n$, in
pristine state stays between 1.0 and 1.2, it increases from 1.4 to
1.7 for 6.15 C/cm$^{2}$, and from 1.7 to 1.9 for 8.5 C/cm$^{2}$
after irradiation. Notably, in the irradiated state, no good fix-exponent
power-law fit was possible above $0.5T_{c}$ and we observed a large
variation of the power-law exponent at lower temperatures prompting
us to explore a different fitting approach. The problem is that, as
we will see from the numerical $t-$matrix analysis ( Fig.\ref{figS2}
and Fig.\ref{figS3} of Appendix A), weak scattering close to Born
limit is not capable of turning $T-$linear behavior to quadratic,
despite the fact that the transition temperature $T_{c}$ is already
substantially suppressed. In other words, $T_{c}$ depends only on
the scattering rate $\Gamma$, but $\Gamma$ is a function of both,
impurity concentration and of the strength of the impurity scattering
potential. For small strength, the suppression of $T_{c}$ is achieved
chiefly due to increasing scatterers concentration. However, the $T^{2}$
behavior seems to be associated with the strong scattering potential,
almost irrespective of the impurity concentration.

\begin{figure}[tb]
\centering \includegraphics[width=7.5cm]{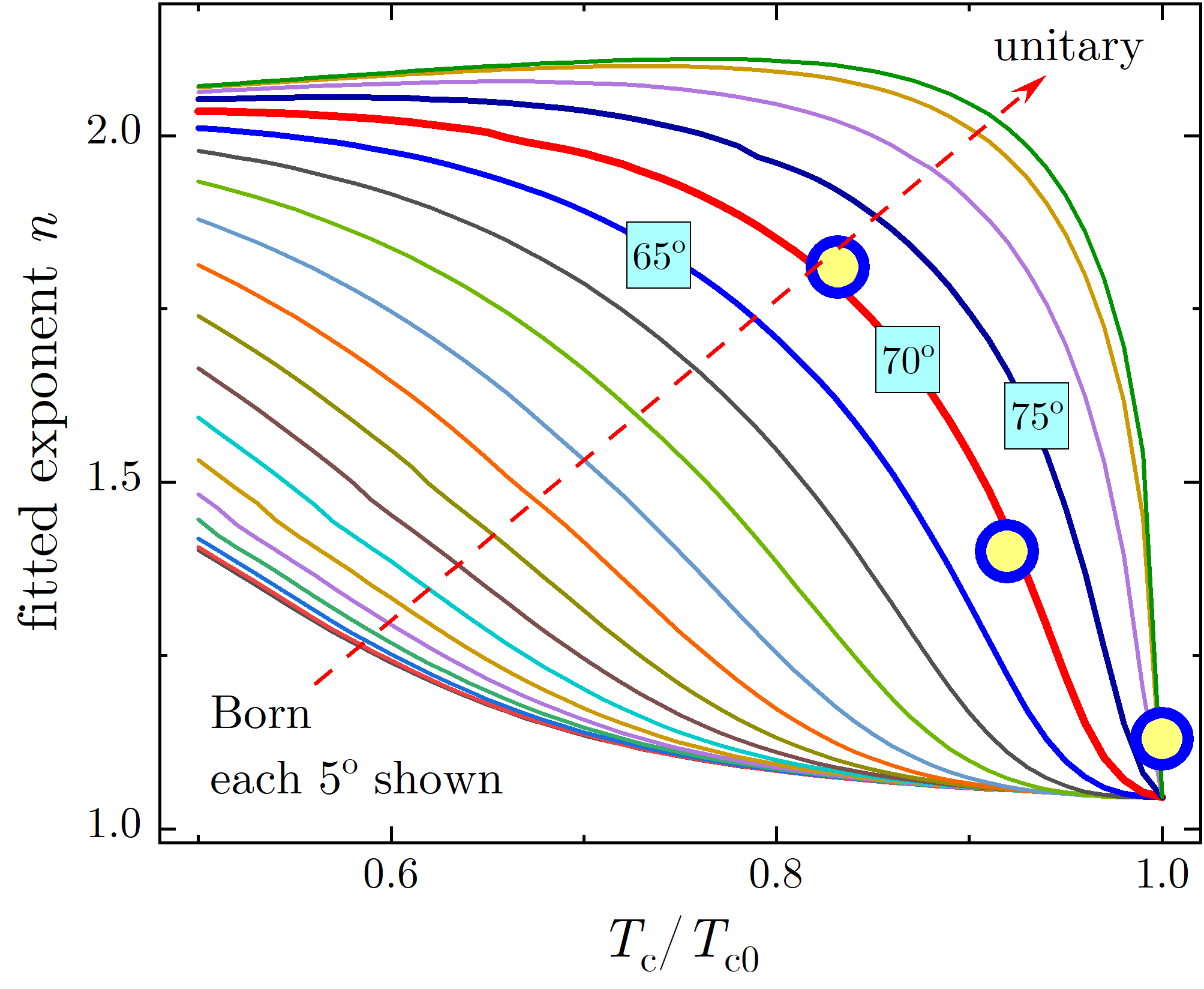} \caption{Lines show the exponent $n$ obtained from the power-law fitting up
to $T_{c}/3$, $\rho_{s}\left(t\right)=1-Rt^{n}$, of the superfluid
density calculated numerically using the $t-$matrix theory. Different
curves correspond to different values of the scattering potential
strength, parameterized by the phase shift $\theta$. The disorder
is presented via the reduced$T_{c}\left(\Gamma\right)/T_{c0}$, also
accessible experimentally. Symbols show the exponents obtained from
the experiment, also fitted up to $T_{c}/3$. The best match is shown
by a bold red line corresponding to $\theta=70^{o}$.}
\label{fig4}
\end{figure}

To analyze the obtained results and compare with the theory, we used
$t-$matrix numerical analysis to calculate the superfluid density,
$\rho_{s}\left(t,\Gamma\right)$. Details of the calculations are
provided in Appendix A. The calculated $\rho_{s}\left(t,\Gamma\right)$
was fitted to the power-law, fitting, $\rho_{s}\left(t\right)=1-Rt^{n}$,
results are shown in Fig.\ref{fig4}. The fit was carried out up to
$T_{c}/3$ for a particular fixed phase shift, $\theta$, and fixed
scattering rate, $\Gamma$, so that the large two-dimensional parameter
space was covered. Each curve in Fig.\ref{fig4} corresponds to a
fixed $\theta$ and the $x-$axis represents the scattering rate via
the reduced suppression of the superconducting transition, $T_{c}\left(\Gamma\right)/T_{c0}\left(\Gamma=0\right)$,
see Fig.\ref{figS1} of the Appendix A. The experimental data, also
taken from the fitting up to $T_{c}/3$, are shown by symbols in Fig.\ref{fig4}.
A good agreement is achieved with $\theta=70^{\circ}$ numerical curve.

To better understand the effect of scattering we now analyze the linear
to quadratic crossover by looking at the polynomial series representation,
which allows comparing the coefficients (unlike the power-law for
which the pre-factor has no particular meaning.) 

\begin{figure}[tb]
\centering \includegraphics[width=7.5cm]{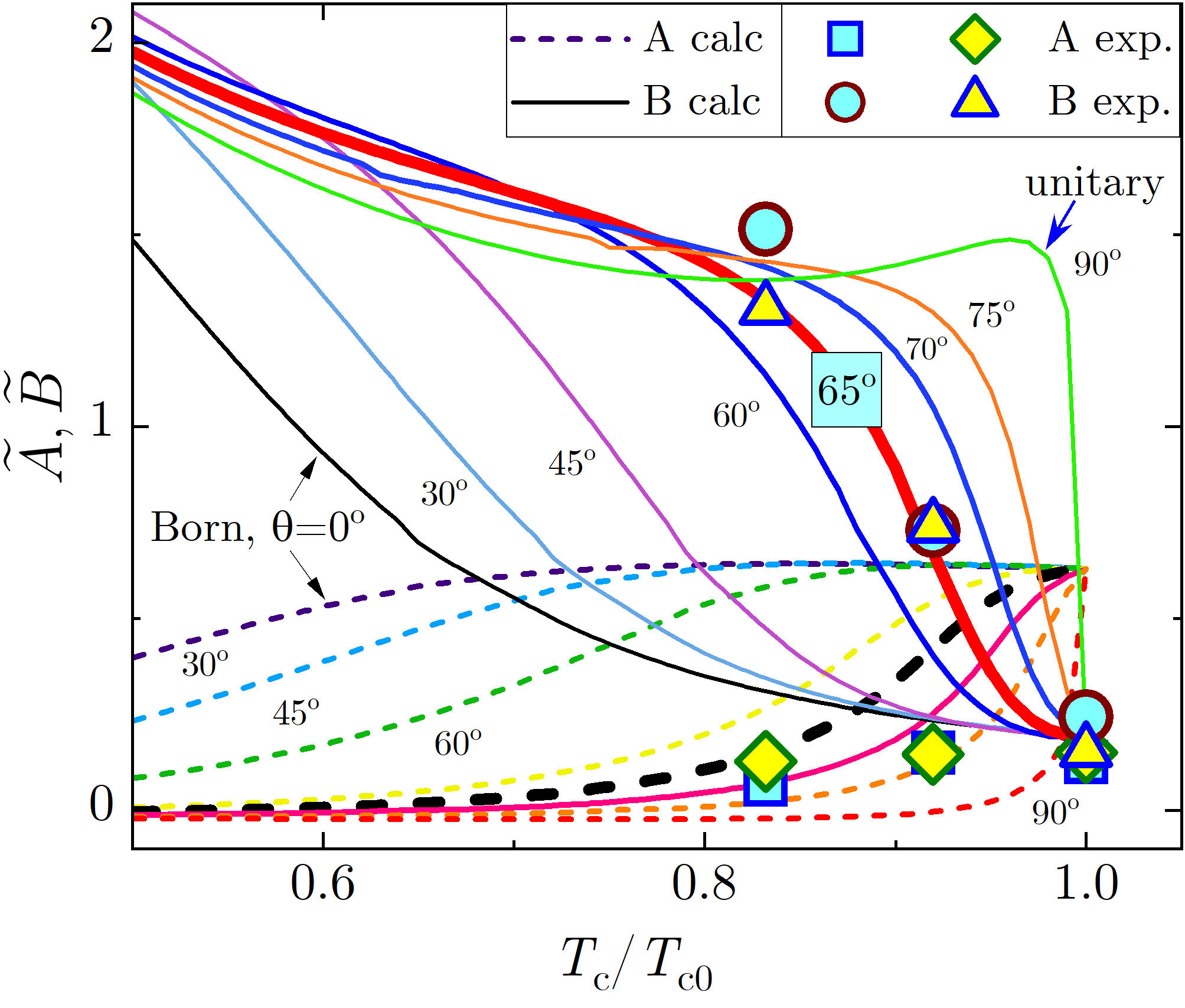} \caption{Coefficients $\widetilde{A}$ and $\widetilde{B}$ obtained from the
fitting of the $t-$matrix superfluid density to the quadratic polynomial,
$\rho_{s}=1-At-Bt^{2}$ and correcting for the change of the magnitude
of $\lambda(0,0)$, namely, $\left[\widetilde{A},\widetilde{B}\right]=\left[A,B\right]\cdot\lambda(0,\Gamma)/\lambda(0,0)$.
This allows using only known quantity, $\lambda(0,0)=150$ nm, for
the experimental $2\Delta\lambda/\lambda(0,0)\approx At+Bt^{2}\approx1-\rho_{s}$.
Dashed lines show $\widetilde{A}$ and solid lines show $\widetilde{B}$
coefficients, respectively. Different curves correspond to fixed values
of the phase shift angle, $\theta$, shown by the labels. The experimental
data are shown by symbols, teal for the upper limit of the polynomial
fitting of $0.3T_{c}$, and yellow for $0.4T_{c}$ upper limit demonstrating
the robust fit. The best match is shown by bold lines corresponding
to $\theta=65^{o}$ phase shift, in a close proximity to the value
of $\theta=70^{o}$ obtained from the power-law fitting.}
\label{fig5} 
\end{figure}

Figure \ref{fig5} shows results of the series polynomial expansion
fit, $\rho_{s}\approx1-At-Bt^{2}$. The experimental data were fitted
to match the same dimensionless coefficients to $2\Delta\lambda/\lambda\left(0,\Gamma\right)\approx At+Bt^{2}\approx1-\rho_{s}$.
The little problem here is that we do not know the value of $\lambda\left(0,\Gamma\right)$
after the irradiation. We therefore used $\lambda(0,0)=150$ nm of
the pristine YBCO to calculate $\Delta\lambda/\lambda(0,0)$ \citep{Bonn1993,Hardy1993PRL,Prozorov2000a}.
Now, multiplying both sides of the above equation by $\lambda(0,\Gamma)/\lambda(0,0)$,
we obtain $2\Delta\lambda/\lambda\left(0,\Gamma\right)\cdot\lambda(0,\Gamma)/\lambda(0,0)=2\Delta\lambda/\lambda\left(0,0\right)\approx\widetilde{A}t+\tilde{B}t^{2}$,
where $\left[\widetilde{A},\widetilde{B}\right]=\left[A,B\right]\cdot\lambda(0,\Gamma)/\lambda(0,0)$.
This way we only use experimental and/or calculated parameters, but
do not guess any numbers. Of course, this assumes that the calculated
$\lambda\left(0,\Gamma\right)$ follows the model, but this is consistent
since the same $t-$matrix model and parameters were used to calculate
it. Figure \ref{fig5} shows the renormalized coefficients $\widetilde{A}$
and $\widetilde{B}$ versus disorder scattering parameterized by $T_{c}/T_{c0}$.
The same numerical $\rho_{s}\left(t,\varGamma/T_{c0}\right)$ curves
were used for the polynomial fitting as for the power-law analysis,
Fig.\ref{fig4}. 

Clearly, the quadratic term (the $B$ coefficient) becomes progressively
more dominant moving towards the unitary limit. However, we find that
the suggested in Ref.\citep{Hirschfeld1993PRB} ``interpolation formula'',
$\Delta\lambda/\lambda\left(0\right)=CT^{2}/\left(T^{*}+T\right)$ has a very narrow range of applicability. For example, by fitting the $t-$matrix numerical results we found that at $T_c/T_{c0}=0.99$, the characteristic temperature is $T^*\approx 0.34T_c$, and at $T_c/T_{c0}=0.98$, $T^*\approx 0.8T_c$, both values already outside the "low-temperature" region of $0.3T_c$. For larger scattering,  $T^*$ practically diverges. Fitting the curves at a fixed $T_c/T_{c0}=0.9$ for different $\theta$ reveals that $T^*$ drops dramatically between $90^{\circ}$ and $70^{\circ}$ and then stays constant at about $T^*\approx 0.025T_c$. 
Therefore, while this approach was very useful for the discussion of the effect of disorder on a $d-$wave superconductor, we should not expect a quantitative agreement of the predicted temperature dependence with the experimental data.

By examining Fig.\ref{fig5}, we can formally assign the experimental
data to $\theta=65^{0}$. These theoretical curves are shown by the
thick lines. For the scattering phase shift of $65^{0}$, we find
the strength of the irradiation induced impurity potential about $70\%$
of the inverse density of states. This is weaker than Schachinger
and Carbotte's estimate \citep{Schachinger2003}. 

\begin{figure}[tb]
\centering \includegraphics[width=8.0cm]{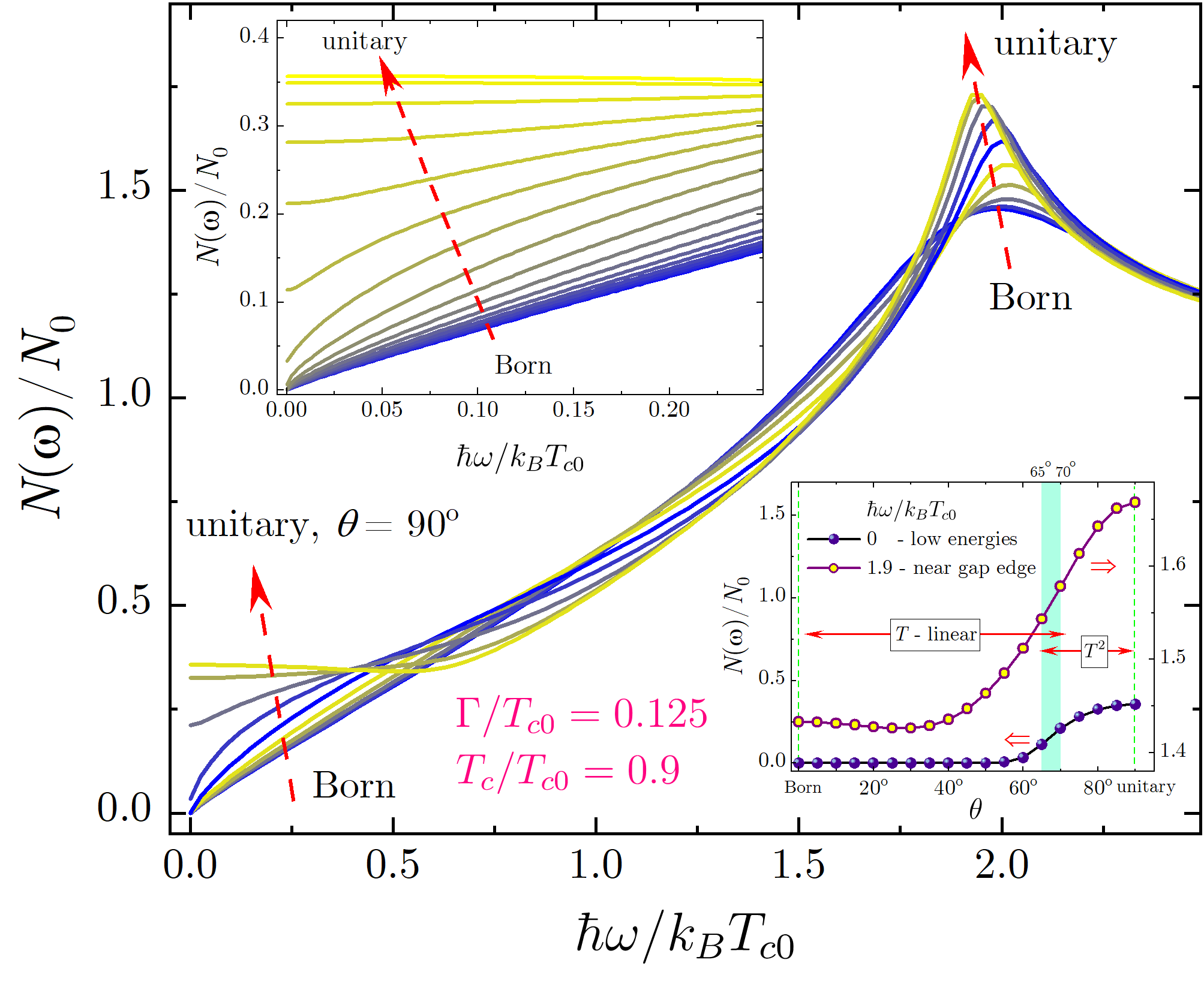} \caption{The normalized density of states spectrum, $N\left(\omega\right)/N_{0}$,
calculated by using a $t-$matrix theory for $\Gamma/T_{c0}=0.125$
corresponding to $T_{c}/T_{c0}=0.9$. The upper inset shows the low-energy
region with a very slow evolution of $N\left(\omega\right)$ for small
$\theta$. The lower inset shows the $\theta-$dependent cuts taken
at the Fermi level, at $\hbar\omega/k_{B}T_{c0}=0$ (left axis) and
near the gap edge, at $\hbar\omega/k_{B}T_{c0}=1.9$ (right axis).
Our experimental estimates of the phase shift between $\theta=65^{\circ}$
and $70^{\circ}$ are shown by the green band. Interestingly that
this is the range of a crossover from $T-$linear and $T^{2}$ behavior.}
\label{fig6}
\end{figure}

To understand our results we analyze the effect of impurity scattering
on the density of states, $N\left(\omega\right)$, for different values
of the phase shift. Here $\hbar\omega$ is the quasiparticle energy
counted from the Fermi energy. Transition temperature, $T_{c}$, depends
only on the scattering rate $\Gamma$, see Fig.\ref{figS1} of Appendix
A, but $\Gamma$ is a function of both, impurity concentration and
of the strength of the impurity scattering potential. We used $t-$matrix
analysis to compute $N\left(\omega\right)$ for a fixed $\Gamma=0.125$,
which correspond to $T_{c}/T_{c0}=0.9.$ Similar results would be
obtained for any other fixed value. The normalized density of states
is, 

\begin{equation}
\frac{N(\omega)}{N_{0}}=\mathrm{Im}\left[\left\langle \frac{\tilde{\omega}}{\sqrt{\Delta_{0}^{2}\cos^{2}\left(2\phi\right)-\tilde{\omega}^{2}}}\right\rangle _{FS}\right]\label{eq:DOS}
\end{equation}
Here $\langle ..\rangle_{FS}$ denotes an average over the Fermi surface,  $N_{0}$ is the normal metal density of states, and $\tilde{\omega}$
is the impurity renormalized energy, see Appendix A for details. The
low-temperature behavior of the superfluid density, hence London penetration
depth, is determined by the low-energy quasiparticle spectrum, $N\left(\omega\rightarrow0\right)$.
Strong and weak scatterers affect the functional form of $N\left(\omega\right)$
in very different ways. Weak scatterers are only able to break the
pairs close to the gap edge, while strong scatterers produce resonant
modes close to zero energy, thus significantly affecting low-temperature
$\rho_{s}\left(t\right)$. Figure \ref{fig6} shows the density of
states spectrum in a full range of energies from zero to above the
gap edge. Each curve is calculated for a fixed phase shift, from Born
to unitary limit, in $10^{\circ}$ steps. Upper inset zooms at the
low energy region showing curves for each $5^{\circ}$. Clearly, low-energy
$N\left(\omega\right)$ barely changes for small values of $\theta\lesssim60^{\circ}$,
but then grow rapidly, and this induces the $T^{2}$ behavior. The
way it happens is visualized in the lower inset in Fig.\ref{fig6},
where $N\left(\omega\right)/N\left(0\right)$ is plotted for two fixed
energies, $\omega=0$, (left $y-$axis) and $\hbar\omega/k_{B}T_{c0}=1.9$
(right $y-$axis) at the peak region near the gap edge. Clearly, the
density of states does not change much until about $50^{\circ}-60^{\circ}$.
Interestingly, this is the region consistent with our experiments.
This re-enforces our conclusion that the observation of linear to
quadratic crossover in the temperature dependence of $\rho\left(t\right)$
is a sensitive indicator of the scattering impurity strength.

We note that our estimates are based on a single-band $d-$wave model
with only one, in-band, scattering channel. However, in the case of
$\text{YBa}_{2}\text{Cu}_{3}\text{O}_{7-\delta}$, there are two bands
crossing the Fermi level centered around the $\mathbf{M}$ point \citep{Borisenko2006_YBCO_ARPES1,YBCO_ARPES2,Dahm2009}.
The presence of multiple band brings additional contribution of the
interband impurity scattering, therefore the single-band estimate
of the impurity potential strength is the upper bound. Experimental
studies of multiband $\text{Ba}_{1-x}\text{K}_{x}\text{Fe}_{2}\text{As}_{2}$
\citep{Cho2016} and $\text{BaFe}_{2}(\text{As}_{1-x}\text{P}_{x})_{2}$
\citep{Mizukami2014}, where electron irradiation was used to introduce
defects, support the intermediate strength of the scattering potential,
but not the unitary limit. 

\section{Conclusions}

Controlled point-like disorder induced by 2.5 MeV low-temperature
electron irradiation was used to suppress superconductivity in $\text{YBa}_{2}\text{Cu}_{3}\text{O}_{7-\delta}$
single crystal with the goal of determining the scattering potential
strength in this high$-T_{c}$ $d-$wave superconductor. The measured
superconducting transition temperature, $T_{c}$, was used as the
measure of the the dimensionless scattering rate, $\Gamma/T_{c0}$.
Normalized superfluid density was obtained from the measured London
penetration depth, $\lambda\left(T\right)$, as well as from the $t-$matrix
theory. By fitting experimental and numerical data to two different
models of the low-temperature behavior we estimated the scattering
potential phase shift is approximately $\theta=65^{\circ}$ to $\theta=70^{\circ}$.
These results find natural explanation when we considered how the
density of states spectra, $N\left(\omega\right)$ are influenced
by weak or strong scatterers. Only the latter affect the low-energy
$N\left(\omega\rightarrow0\right),$which determines the low-temperature
behavior of $\rho_{s}\left(t\right)$. Our results provide certain
boundary conditions for the microscopic investigation of the pairing
interactions and scattering in high$-T_{c}$ cuprates for which the
knowledge of the scattering phase shift is vital.

\section*{Acknowledgments}

We thank Peter Hirschfeld and David Broun for useful discussions.
This work was supported by the U.S. Department of Energy (DOE), Office
of Science, Basic Energy Sciences, Materials Science and Engineering
Division. Ames Laboratory is operated for the U.S. DOE by Iowa State
University under contract DE-AC02-07CH11358. VM is supported by NSFC
grant 11674278 and by the priority program of the Chinese Academy
of Sciences grant No. XDB28000000. We thank the SIRIUS team, O. Cavani,
B. Boizot, V. Metayer, and J. Losco, for running electron irradiation
at École Polytechnique supported by the EMIR (Réseau national \'{d}accélérateurs
pour les Etudes des Matériaux sous Irradiation) network user proposal.


%

\appendix

\section{$t-$matrix approximation for a single-band $d-$wave superconductor
with non-magnetic disorder scattering}

This appendix provides some technical details of the $t-$matrix numerical
analysis used in this paper.

\begin{figure}[tb]
\includegraphics[width=8cm]{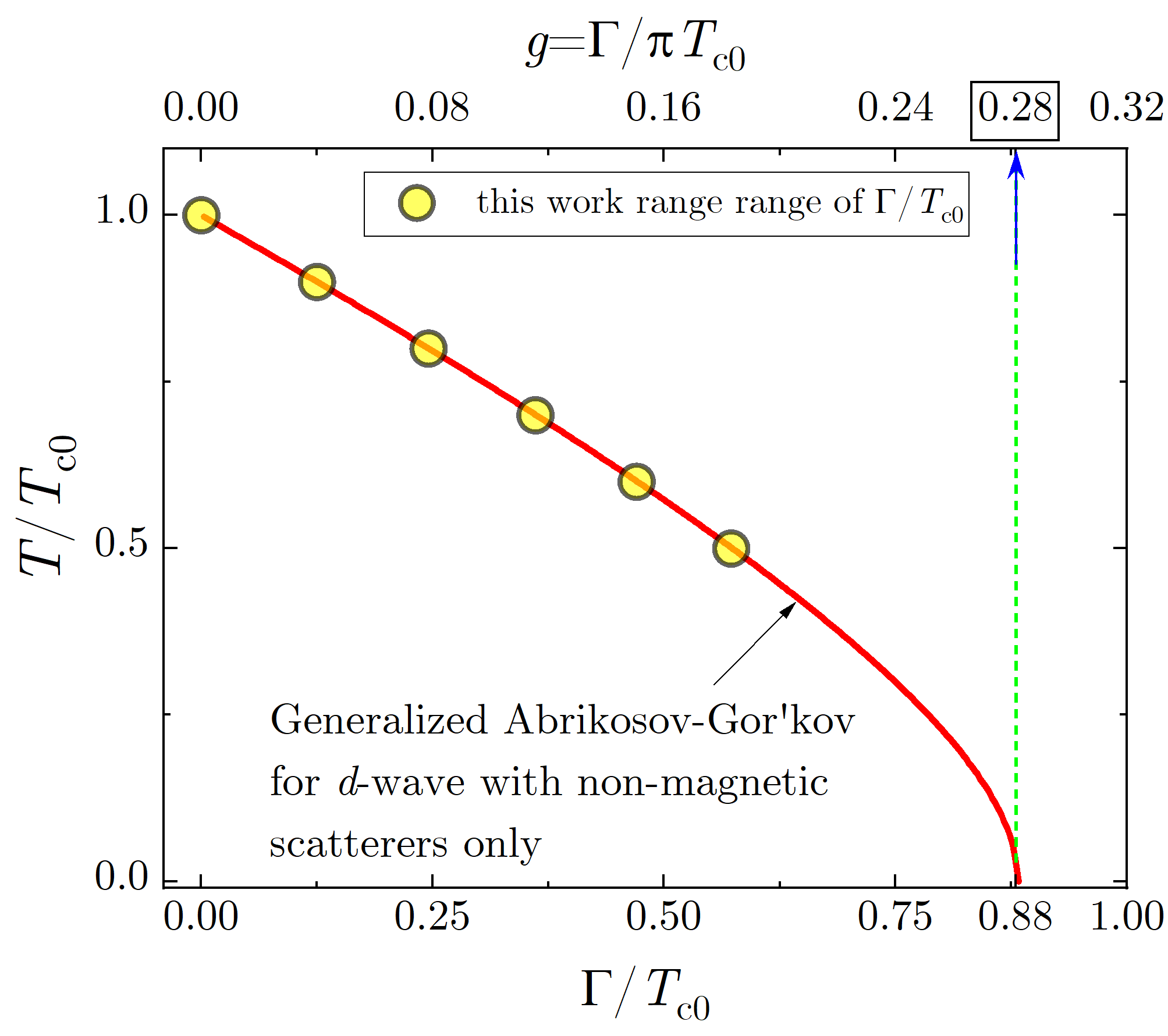} 
\caption{Universal ``Abrikosov-Gor'kov'' - type dependence of the normalized
superconducting transition temperature, $T_{c}/T_{c0}$, on the dimensionless
scattering rate, $\Gamma/T_{c0}$. Bottom and top axes follow slightly
different definitions used in the literature. Solid line shows the
numerical solution of the Abrikosov-Gor'kov equation extended to arbitrary
angular dependence of the order parameter, in this particular case
to nonmagnetic impurities in a $d-$wave superconductor. Symbols show
the range covered by the numerical $t-$matrix calculations presented
in this paper.}
\label{figS1}
\end{figure}

The $t-$matrix approximation for the impurity scattering is based
on the summation of all the single site impurity scattering diagrams
\citep{PJH_Wolfle1988}, hence, it is exact in the low impurity concentration
limit. In general, the impurity dressed Green's function for a $d$-wave
superconductor reads,

\begin{eqnarray}
\mathbf{G}(\mathbf{k},i\omega_{n})=-\frac{i\tilde{\omega}_{n}\tau_{0}+\xi_{\mathbf{k}}\tau_{3}+\Delta_{\mathbf{k}}\tau_{1}}{\tilde{\omega}_{n}^{2}+\xi_{\mathbf{k}}^{2}+\Delta_{\mathbf{k}}^{2}},\label{Eq:FullG}
\end{eqnarray}

\noindent where $\xi_{\mathbf{k}}$ is the electronic dispersion, $\Delta_{\mathbf{k}}$
is the momentum superconducting gap, and $\omega_{n}$ is the fermionic
Matsubara frequency. The impurity renormalized Matsubara frequency
id denoted by $\tilde{\omega}_{n}$, which is 
\begin{eqnarray}
\tilde{\omega}_{n}=\omega_{n}+\frac{n_{imp}}{\pi N_{0}}\frac{g_{0}}{\cot^{2}\theta_{s}+g_{0}^{2}},
\end{eqnarray}
here $n_{imp}$ is the impurity concentration, $N_{0}$ is the density
of states at the Fermi energy, $g_{0}$ is, 
\begin{equation}
g_{0}=\int_{0}^{2\pi}\frac{d\phi}{2\pi}\frac{\tilde{\omega}_{n}}{\sqrt{\tilde{\omega}_{n}^{2}+\Delta_{\phi}^{2}}},
\end{equation}
and $\theta_{s}$ is the $s$-wave scattering phase shift, defined
as, 
\begin{eqnarray}
\tan\theta_{s}=\pi N_{0}V_{imp}.
\end{eqnarray}

\noindent In the unitary limit, the $s$-wave scattering phase shift
becomes $\pi/2$ as the impurity potential $V_{imp}$ goes to $\infty$.
Note, we assume an isotropic Fermi surface and restrict the momentum
dependence of the gap to the Fermi surface ($\Delta_{\mathbf{k}}=\Delta_{0}\cos2\phi$),
this is a reasonable approximation in the low temperature limit.

\begin{figure}[tb]
\includegraphics[width=8cm]{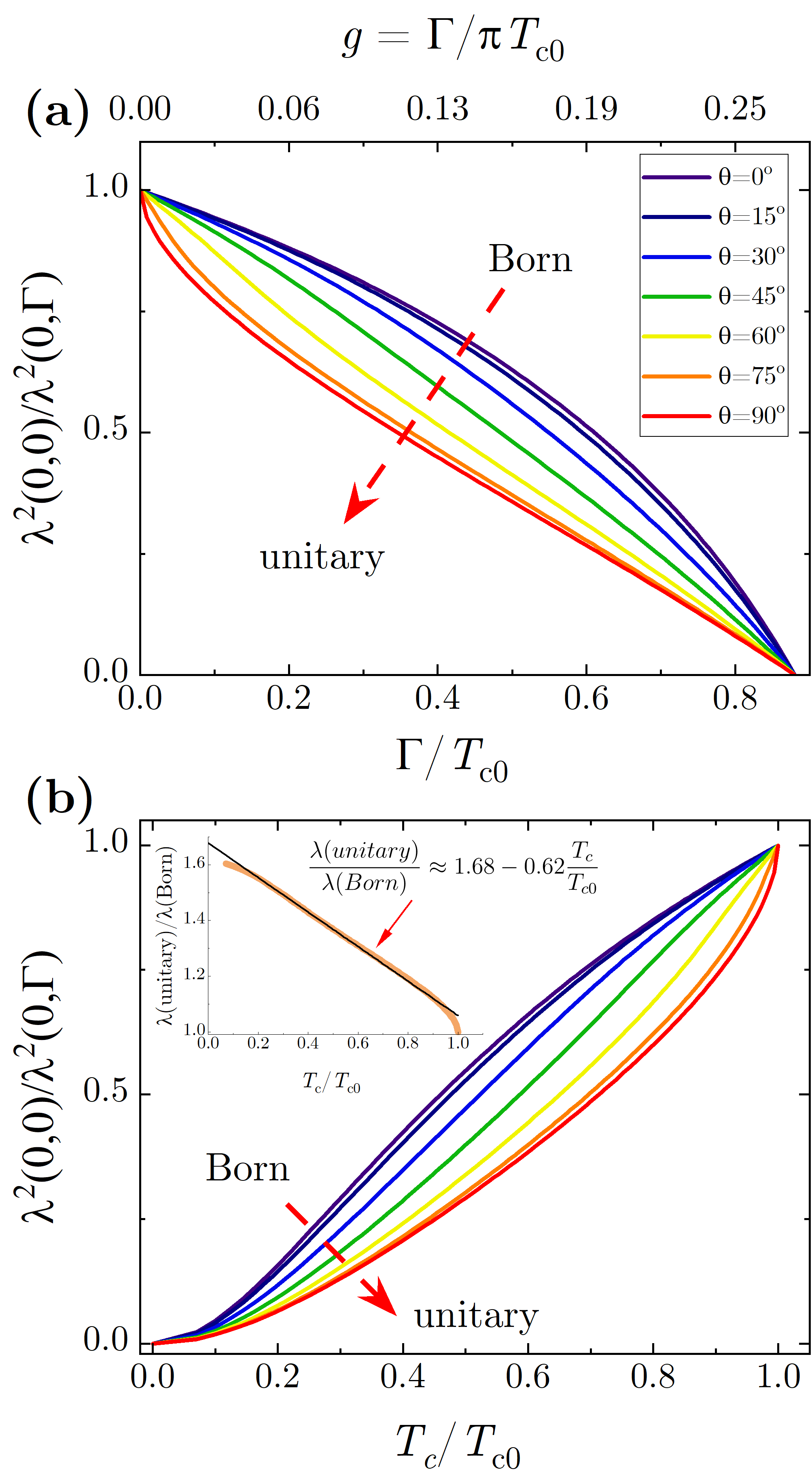} 
\caption{Suppression of the superfluid density at $T=0$ by nonmagnetic impurities
for \textbf{(a)} different strengths of the scattering potential varying
between Born and unitary limits, and \textbf{(b)} as function of the
measurable quantity, $T_{c}/T_{c0}$. Each curve corresponding to
the fixed angle $\theta$ indicated in the legend. Panel \textbf{(a)}
uses dimensionless scattering rates as top and bottom $x-$axes. For
a fixed scattering rate, London penetration depth, expectedly, increases
the most in the unitary limit, however it remains finite and the total
change between two limits is less than a factor of two. This is shown
in the inset in panel \textbf{(b)}. A simple practical formula allowing
estimation of the amplitude of the unitary limit is given, $\lambda(\text{unitary})$/$\lambda\left(\text{Born}\right)\approx1.68-1.62\left(T_{c}/T_{c0}\right)$.}
\label{figS2}
\end{figure}

The impurity dressed Green's function is used to calculate the gap,
\begin{eqnarray}
\Delta_{0} & = & 2\pi TV_{d}N_{0}\sum_{\omega_{n}>0}^{\Omega_{c}}\int_{0}^{2\pi}\frac{d\phi}{2\pi}\frac{\Delta_{0}\cos^{2}2\phi}{\sqrt{\tilde{\omega}_{n}^{2}+\Delta_{0}^{2}\cos^{2}2\phi}}.
\end{eqnarray}
Here we choose a simple separable pairing interaction $V_{pair}=-V_{d}\cos2\phi\cos2\phi'$,
and $\Omega_{c}$ is the pairing cutoff energy scale. A negative value
of $V_{d}$ means attractive interaction for $d$-wave superconductivity.
The transition temperature is determined by, 
\begin{eqnarray}
\log\left(\frac{T_{c}}{T_{c0}}\right) & = & \Psi\left(\frac{1}{2}\right)-\Psi\left(\frac{1}{2}+\frac{\Gamma}{2\pi T_{c}}\right),
\end{eqnarray}
where $T_{c0}$ is the clean limit $T_{c}$, $\Psi$ is the digamma
function, and $\Gamma$ is the pair-breaking energy-scale, which is
the normal state single particle scattering rate, and within the t-matrix
approximation reads, 
\begin{eqnarray}
\Gamma & = & \frac{n_{imp}}{\pi N_{0}}\frac{1}{\cot^{2}\theta_{s}+1}.
\end{eqnarray}
Note, for a single band $d$-wave superconductor, there is a universal
energy scale that determines $T_{c}$ suppression. In the Born limit,
$\Gamma$ is $n_{imp}\pi N_{0}V_{imp}^{2}$. The variation of $T_{c}$
as a function of this universal energy scale is shown in Fig.\ref{figS1}.
The critical value of $\Gamma$ sufficient to destroy superconductivity
is $\pi\exp^{-\gamma}/2T_{c0}$, where $\gamma\approx0.577$ is the
Euler-Mascheroni constant. Once we calculate the impurity dressed
Green's function, we can calculate the superfluid density, 
\begin{eqnarray}
\rho_{s}=\frac{n_{s}}{n} & = & 2\pi T\sum_{\omega_{n}>0}\int_{0}^{2\pi}\frac{d\phi}{2\pi}\frac{\Delta_{0}^{2}\cos^{2}2\phi}{\left(\tilde{\omega}_{n}^{2}+\Delta_{0}^{2}\cos^{2}2\phi\right)^{3/2}},
\end{eqnarray}
where $n$ is the normal fluid density.  The magnetic penetration depth
can be expressed as,  $\lambda (T,\Gamma)/\lambda(0,0)=1/\sqrt{\rho_{s}}$. Apart from magnetic penetration depth, another quantity of interest is the density of state (see Eq. \eqref{eq:DOS}). For the density of state, we perform the analytical continuation of the Matsubara frequency to the upper half the plane ($i\omega_n \rightarrow \omega + i 0^+$) to obtain the renormalized energy $\omega$. After analytic continuation, the equation for the renormalized energy reads,
\begin{equation}
  \tilde{\omega} = \omega + i 0^+ + \frac{n_{imp}}{\pi N_0}   \frac{g_\omega}{\cot^2 \theta_s - g_\omega^2},
\end{equation}
where $g_\omega$ is,
\begin{eqnarray}
g_\omega = \int_0^{2\pi} \frac{d\phi}{2\pi} \frac{\tilde{\omega}}{\sqrt{\Delta_0^2 \cos^2 2\phi-\tilde{\omega}^2}}.
\end{eqnarray}
Note $\tilde{\omega}$ is a complex quantity.

\begin{figure}[tb]
\includegraphics[width=8.5cm]{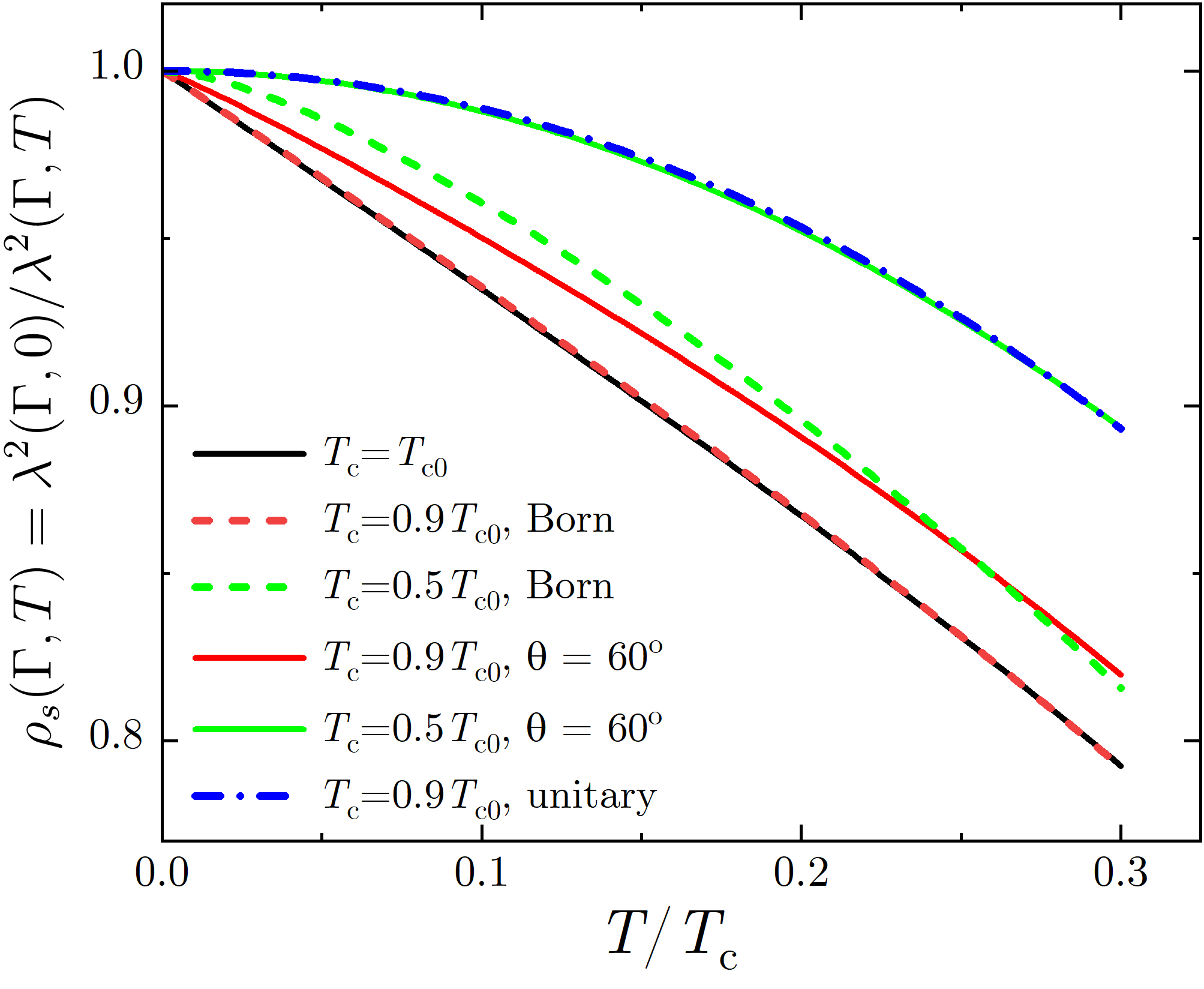} \caption{Normalized superfluid density as function of a reduced temperature
in Born, unitary and intermediate scattering rate. For the same rate
of $T_{c}$ suppression, the curve is practically unchanged in the
Born limit and shows quadratic behavior in unitary limit even for
a moderate suppression.}
\label{figS3}
\end{figure}

Figure \ref{figS2} shows the suppression of the superfluid density
at $T=0$ by nonmagnetic impurities for \textbf{(a)} different strengths
of the scattering potential varying between Born and unitary limits,
and \textbf{(b)} as function of the measurable quantity, $T_{c}/T_{c0}$.
Each curve corresponding to the fixed angle $\theta$ indicated in
the legend. Panel \textbf{(a)} uses two slightly different definitions
of the dimensionless scattering rates as top and bottom $x-$axes.
For a fixed scattering rate, London penetration depth, expectedly,
increases the most in the unitary limit, however it remains finite
and the ratio between the two limits is less than a factor of two.
This is shown in the inset in panel \textbf{(b)}. A simple practical
formula allowing estimation of the amplitude of the unitary limit
is given, $\lambda(\text{unitary})$/$\lambda\left(\text{Born}\right)\approx1.68-1.62\left(T_{c}/T_{c0}\right)$.
Note, the variation of superfluid density is not universal like $T_{c}\left(\Gamma\right)$.
This also reflects in the temperature dependence of the superfluid
density in the Fig. \ref{figS3}, where the unitary or stronger impurities
brings the quadratic term very quickly for a very little $T_{c}$
suppression.

\section{Electron irradiation}

Point-like disorder was introduced at the SIRIUS facility in the Laboratoire
des Solides Irradiés at École Polytechnique in Palaiseau, France.
Electrons accelerated in a pelletron-type linear accelerator to 2.5
MeV displace ions creating vacancy - interstitial pairs (known as
``Frenkel pairs'') \citep{Damask1963,THOMPSON1969}. The acquired
irradiation dose is determined by measuring the total charge accumulated
by a Faraday cage located behind the sample. The sample is held in
liquid hydrogen at around 20 K needed not only to remove significant
amount of heat produced by sub-relativistic electrons, but also to
prevent immediate recombination and migration of the produced atomic
defects. On warming to room temperature the interstitials that have
lower barrier of diffusion, migrate to various sinks (dislocations,
twin boundaries, surfaces) faster, leaving the metastable population
of vacancies. The achieved level of disorder induced by the irradiation
is gauged by the change of resistivity. Detailed studies of YBCO samples
from the same source is found elsewhere \citep{Rullier-Albenque2003PRL_YBCO_e-irr,Rullier_Albenque2000EPL_YBCO_e-irr}.

\begin{figure}[tb]
\includegraphics[width=8.5cm]{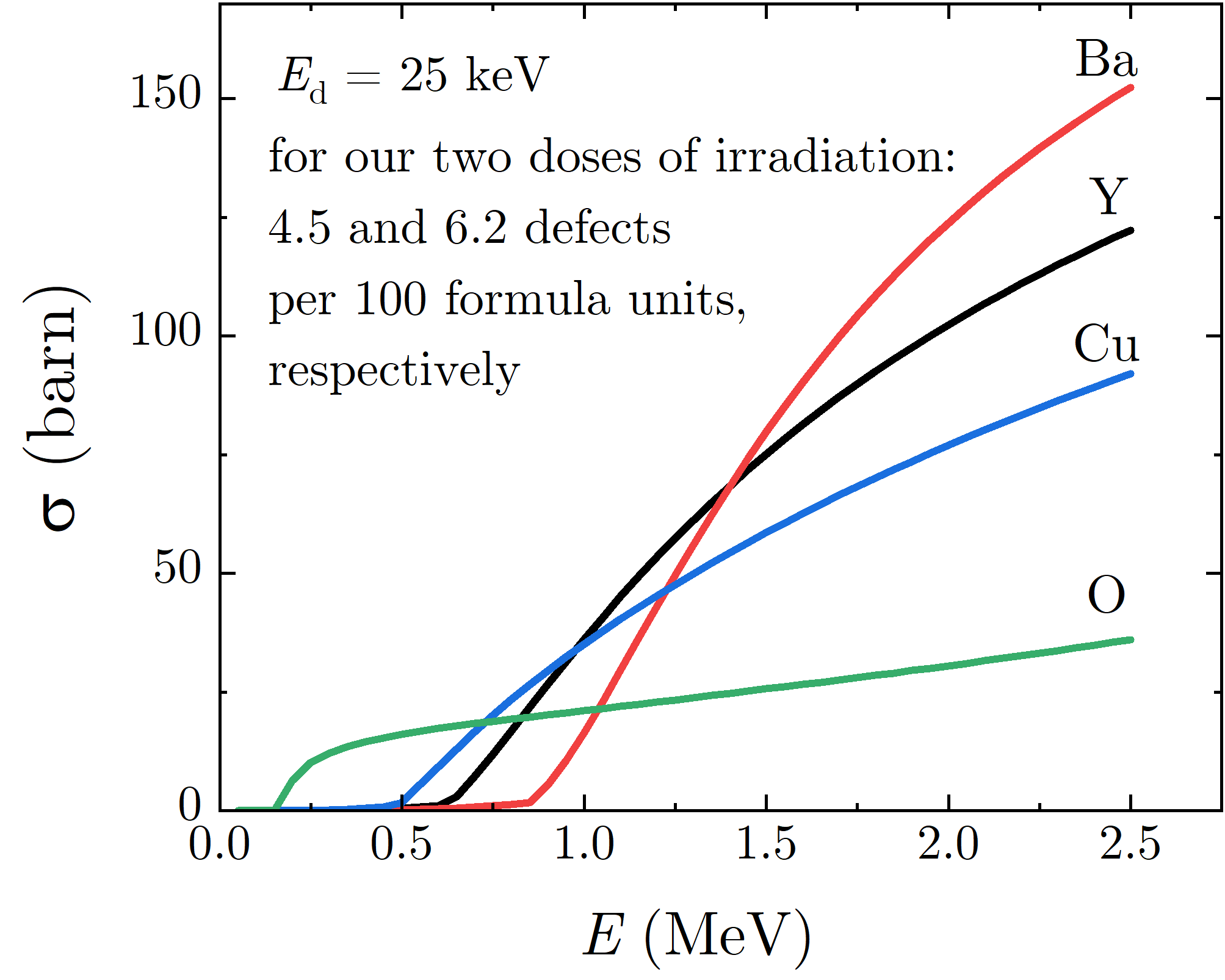} \caption{Partial cross-section of atomic defects creation by electron irradiation
of indicated energy. At our beam energy of 2.5 MeV the estimated total
cross-section of any atom gives 4.5 and 6.3 defects per hundred formula
units, assuming equal knock-out energy of 25 eV, typical for this
kind of material \citep{Damask1963,THOMPSON1969}. The cross-sections
are calculated using proprietary code SECTE developed at École Polytechnique
specifically to describe electron irradiation experiments used in
this paper.}
\label{figS4} 
\end{figure}

Ion-resolved cross-sections were calculated using SECTE software package
developed at École Polytechnique, France specifically for the electron
irradiation. Among different projectiles, electrons are best to produce
point-like defects due to their small rest mass. As shown in Fig.\ref{figS4},
at our energy of 2.5 MeV, all YBCO ions are active and estimated density
of defects of any kind is about 5 defects per 100 formula units, which
means that the defects are well separated and do not alter the material
itself.

\end{document}